\documentstyle [12pt] {article}
\begin{document}

\pagestyle{empty}
\rightline{\vbox{
\halign{&#\hfil\cr
&ANL-HEP-PR-92-63\cr
&NUHEP-TH-92-16\cr
&August 1992\cr}}}
\bigskip
\bigskip
\bigskip
{\Large\bf
	\centerline{P-Wave Charmonium Production}
	\centerline{in $B$-Meson Decays}}
\bigskip
\normalsize
\centerline{Geoffrey T. Bodwin}
\centerline{\sl High Energy Physics Division, Argonne National Laboratory,
    Argonne, IL 60439}
\bigskip

\centerline{Eric Braaten and Tzu Chiang Yuan}
\centerline{\sl Department of Physics and Astronomy, Northwestern University,
    Evanston, IL 60208}
\bigskip

\centerline{G. Peter Lepage}
\centerline{\sl Newman Laboratory of Nuclear Studies, Cornell University,
    Ithaca, NY 14853}
\bigskip

\begin{abstract}
We calculate the decay rates of $B$ mesons into P-wave
charmonium states using new
factorization formulas that are valid to leading order in the
relative velocity of the charmed quark and antiquark and to all orders in the
running coupling constant of QCD.  We express the production rates for all four
P states in terms of two nonperturbative parameters,
the derivative of the wavefunction at the origin
and another parameter related to
the probability for a charmed-quark-antiquark pair in a
color-octet S-wave state to radiate a soft gluon
and form a P-wave bound state.  Using existing data on
$B$ meson decays into $\chi_{c1}$ to estimate the color-octet parameter,
we find that the color-octet mechanism may
account for a significant fraction of the $\chi_{c1}$ production rate
and that $B$ mesons should decay into $\chi_{c2}$ at a similar rate.
\end{abstract}
\vfill\eject\pagestyle{plain}\setcounter{page}{1}

The production rate of quarkonium states in various high energy physics
processes can provide valuable insight not only into the interactions between
a heavy quark and antiquark, but also into the elementary
processes that produce the $Q {\bar Q}$ pair.
The spin-1 S-wave resonances, like the $J/\psi$ of charmonium,  are of
special experimental significance because they have extremely clean
signatures through their leptonic decay modes.
Because a significant fraction
of the $\psi$'s come from the decays of the P-wave $\chi_{cJ}$ states,
an understanding of the production of P-wave resonances is necessary
in order to understand inclusive $\psi$ production.
The P states are also important in their own right
because they probe a qualitatively different
aspect of the $Q {\bar Q}$ production process.
While the S states probe only the production
at short distances of a $Q {\bar Q}$ pair in a color-singlet state,
the P states, as we shall show, also probe the production of a $Q {\bar Q}$
pair
in a color-octet state.

One of the simplest production processes for charmonium states
is the decay of a $B$ meson or baryon.
For $B^-$, ${\bar B}^0$, ${\bar B}_s$, and $\Lambda_b$
(but not for ${\bar B}_c$),
the $c$ and the ${\bar c}$ that form the charmonium bound state must both be
produced by the decay or annihilation of the $b$ quark.  Since this
is a short distance process which occurs on a length scale
of order $1/M_b$, where $M_b$ is the mass of the $b$ quark,
it should be possible to apply perturbative QCD to calculate
the inclusive decay rate into a particular charmonium state.
If we neglect contributions that are suppressed by powers of $\Lambda/M_b$,
where $\Lambda$ is a typical momentum scale for light quarks,
the decay rate of a $B$ hadron is given by the
decay rate of the $b$ quark, with the light antiquark in the $B$ meson
and the light quarks in the $B$ baryon treated as noninteracting spectators.
In the case of the hadronic and the semileptonic decay rates,
the leading corrections are suppressed by
two powers of $\Lambda/M_b$ (Ref. \cite{bigi}).  We expect
perturbative QCD calculations to yield predictions for the
inclusive decay rates of $B$ hadrons into
charmonium states that are comparable in accuracy to the predictions
for their semileptonic decay rates.

Most previous calculations of the rate for charmonium production
in $B$ meson decays \cite{mbw,dt,knr,chjc} have
been carried out under the assumption that the production
mechanism is the decay at short distances of a
$b$ quark into a color-singlet $c {\bar c}$ pair
plus other quarks and gluons, with the $c$ and  ${\bar c}$ having almost
equal momenta and residing in the appropriate angular-momentum state.
We will refer to this as the ``color-singlet mechanism''.
It was assumed that the only nonperturbative input required in
the calculation is the nonrelativistic $c {\bar c}$
wave function (for S states) or its derivative
(for P states) at the origin.
In this paper we point out that this assumption fails for P states,
the failure being signaled by the presence
of infrared divergences in the QCD radiative corrections.
The divergences appear because
there is a second production mechanism for P states
that also contributes at leading order in $\alpha_s$, namely
the decay at short distances of the $b$ quark into a color-octet
$c {\bar c}$ pair in an S state, plus other partons.
We will call this the ``color-octet mechanism''.  In addition to the
derivative of the nonrelativistic wavefunction at the origin,
calculations of P-wave production
require a second nonperturbative input parameter,
the probability for the color-octet $c {\bar c}$ pair
to radiate a soft gluon and form a color-singlet P-wave bound state.
The two production mechanisms are summarized below by
factorization formulas which are accurate to leading order in $v^2$,
the square of the relative velocity of the charmed quark and antiquark,
and to all orders in the strong coupling constant $\alpha_s(M_b)$.
The hard subprocess rates appearing in these formulas are calculated
to leading order in the QCD coupling constant.  Estimates of
the nonperturbative parameter associated with the color-octet mechanism
are obtained from experimental data on the production rate of $\chi_c$ states
in $B$ meson decays.  We find that the color-octet mechanism may
account for a significant fraction of the observed decay rate
into $\chi_{c1}$ and that $B$ mesons should decay
into $\chi_{c2}$ at a similar rate.

The color-singlet and color-octet mechanisms for the production of
P-wave quarkonium states have analogs in the decays of these states.
For S-wave resonances, the electromagnetic and light-hadronic decays
proceed through the annihilation at short distances
of the heavy quark and antiquark in  a color-singlet S-wave state.
The decay rates are then proportional to the square of $R_S(0)$,
the nonrelativistic wavefunction at the origin, with coefficients
that can be calculated perturbatively as a series in $\alpha_s(M_Q)$,
where $M_Q$ is the heavy quark mass.
In the case of P-waves, one might expect that the decay should
also proceed through the annihilation at short distances
of the $Q {\bar Q}$ pair in  a color-singlet P-wave state.
If this were the case, the decay rates would be calculable in terms of a single
nonperturbative factor $R_P'(0)$,
the derivative of the nonrelativistic wavefunction at the origin.
While this expectation holds true for decay rates into two photons,
explicit calculations of the light hadronic decay rates reveal
infrared divergences \cite{barb}.
It has recently been shown that these infrared divergences can be
systematically factored into a second nonperturbative parameter,
which is proportional to the probability for the quark and antiquark
to be in a color-octet S-wave state at the origin \cite{bbla}.
These results can be summarized by rigorous factorization formulas
for the decay rates of quarkonium states into light hadronic or
electromagnetic final states \cite{bbla,bblb}.
The factorization formulas are valid
to leading order in $v^2$, where $v$ is the typical velocity
of the heavy quark, and to all orders in the QCD coupling constant
$\alpha_s(M_Q)$.

The inclusive production rates for charmonium states
also obey factorization formulas.
To leading order in $v^2$, where $v$ is the typical velocity of the
charmed quark relative to the charmed antiquark,
and to all orders in $\alpha_s(M_b)$,
the inclusive decay rate into $\psi$ satisfies
the simple factorization formula that was assumed in previous
work: \cite{mbw,dt,knr,chjc}
\begin{equation} {
\Gamma \left( b \rightarrow \psi + X \right) \; = \;
G_1 \; {\widehat \Gamma}_1
\left( b \rightarrow c {\bar c} (^3S_1) + X \right) \; ,
} \label{factpsi}
\end{equation}
where ${\widehat \Gamma}_1$ is the hard subprocess rate for producing a
color-singlet $c {\bar c}$ pair with equal momenta and in the
appropriate angular-momentum state $^{2S+1}L_J = \; ^3S_1$.
It can be calculated perturbatively as a series
in the QCD running coupling constant $\alpha_s(M_b)$.
The nonperturbative parameter $G_1$ in
(\ref{factpsi}) is proportional to the probability for the
$c {\bar c}$ pair to form a bound state and is related to the
nonrelativistic wavefunction at the origin $R_S(0)$:
\begin{equation} {
G_1\; \approx \;
{3 \over 2 \pi} {|R_S(0)|^2 \over M_c^2} \; .
} \label{gone}
\end{equation}
A phenomenological determination that makes use of the electronic
decay rate of the $\psi$ gives $G_1 \approx 108$ MeV.  It can also be given a
rigorous nonperturbative definition, so that it can be
measured in lattice simulations of QCD \cite{bblb}.
The production rate of the radial excitation $\psi'$ obeys the same
factorization formula (\ref{factpsi}), but with a parameter $G_1$
that is smaller by the ratio of the electronic decay rates of
$\psi$ and $\psi'$:  $G_1 \approx 43$ MeV.

In previous work \cite{knr}, the decays of $b$ quarks into P-wave
charmonium states were assumed to satisfy simple
one-term factorization formulas like (\ref{factpsi}).
Because of the color-octet production mechanism, such formulas are
incomplete, and their breakdown is signaled by the appearance of
infrared divergences in order $\alpha_s$.
The correct factorization formulas for P-wave production
rates have two terms:
\begin{equation}{
\Gamma \left( b \rightarrow h_c + X \right) \; = \;
H_1 \; {\widehat \Gamma}_1
\left( b \rightarrow c {\bar c} (^1P_1) + X, \mu\right) \;
+ \;  3 \; H_8'(\mu) \; {\widehat \Gamma}_8
\left( b \rightarrow c {\bar c} (^1S_0) + X \right) \; ,
} \label{facth}
\end{equation}
\begin{equation}{
\Gamma \left( b \rightarrow \chi_{cJ} + X \right) \; = \;
H_1 \; {\widehat \Gamma}_1
\left( b \rightarrow c {\bar c} (^3P_J) + X, \mu \right) \;
+ \;  (2J + 1) \; H_8'(\mu) \; {\widehat \Gamma}_8
\left( b \rightarrow c {\bar c} (^3S_1) + X \right) \; .
} \label{factchi}
\end{equation}
The factors ${\widehat \Gamma}_1$ and ${\widehat \Gamma}_8$
are hard subprocess rates for the decay of the
$b$ quark into a $c {\bar c}$ pair in the appropriate angular-momentum state
with vanishing relative momentum, the $c {\bar c}$
being in a color-singlet state for ${\widehat \Gamma}_1$
and in a color-octet state for ${\widehat \Gamma}_8$.
They can be calculated perturbatively as series in $\alpha_s(M_b)$,
with coefficients that are free of infrared divergences.
Note that in (\ref{factchi}) the only dependence on
the total angular momentum quantum number $J$ in the second term on
the right side lies in the coefficient $2J + 1$.
The nonperturbative parameters $H_1$ and $H_8'$
are proportional to the probabilities for
a $c {\bar c}$ pair in a color-singlet P-wave
and a color-octet S-wave state, respectively,
to fragment into a color-singlet P-wave bound state.
They can be given rigorous nonperturbative definitions \cite{bblb}
in terms of matrix elements in nonrelativistic QCD.
The parameter $H_1$ is directly related to the
nonrelativistic wavefunction for the heavy quark and antiquark:
\begin{equation} {
H_1 \; \approx \; {9\over 2\pi} {|R^\prime_P(0)|^2 \over M_c^4} \;.
} \label{hsinglet}
\end{equation}
Its value has been determined phenomenologically from the light hadronic decay
rates of the $\chi_{c1}$ and $\chi_{c2}$ to be $H_1 \approx 15$ MeV
\cite{bbla}.
There is no rigorous perturbative expression for $H_8'$ in terms of
the wavefunction $R_P(r)$,
since a $c {\bar c}$ pair in a color-octet S-wave state can make a
transition to a color-singlet P-wave state through the
radiation of a soft gluon,
which is a nonperturbative process.  The parameter $H_8'$ is not
related in any simple way to the analogous parameter $H_8$
that appears in P-wave decays \cite{bbla}, so
it cannot be determined from data on the decays of P states.
A phenomenological determination of $H_8'$ can come only
from a production process.
Below we will use experimental data on $\chi_c$ production
in $B$ meson decays to obtain a rough determination of $H_8'$.

In the factorization formulas (\ref{facth}) and (\ref{factchi}),
$H_8'$ and ${\widehat \Gamma}_1$ depend on an arbitrary
factorization scale $\mu$ in such a way that the complete decay rate is
independent of $\mu$.
In order to avoid large logarithms of $\mu/M_b$ in
${\widehat \Gamma}_1$, the factorization scale $\mu$ should be chosen to
be on the order of $M_b$.
The scale dependence of $H_8'(\mu)$ is governed by a
renormalization group equation, which to leading order in
$\alpha_s(\mu)$ is \cite{bblb}
\begin{equation} {
\mu {d \over d \mu} H_8'(\mu) \;\approx\;
{16 \over 27 \pi} \alpha_s(\mu) \; H_1 \;.
} \label{rgeq}
\end{equation}
The matrix element $H_1$ is independent of the factorization scale $\mu$.
The solution to the renormalization group equation is
therefore elementary.
For example, for $\mu < M_c$, the solution is
\begin{equation} {
 H_8'(M_b) \;=\; H_8'(\mu) \;+\; \left[ {16 \over 27 \beta_3}
\ln  \left( {\alpha_s(\mu) \over \alpha_s(M_c)} \right)
\;+\; {16 \over 27 \beta_4}
\ln  \left( {\alpha_s(M_c) \over \alpha_s(M_b)} \right) \right] \, H_1 \; ,
} \label{rgsol}
\end{equation}
where $\beta_n = (33-2n)/6$ is the first coefficient of the
QCD beta function for $n$ flavors of massless quarks.
In the limit in which $M_b$ is very large,
the contribution to $H_8'$ from the perturbative evolution dominates,
and one can estimate $H_8'(M_b)$ by setting $\alpha_s(\mu) \sim 1$
and neglecting the constant $H_8'(\mu)$ in (\ref{rgsol}).
Taking $\alpha_s(M_b) \approx 0.20$ and $\alpha_s(M_c) \approx 0.31$,
we obtain $H_8'(M_b) \approx 3$ MeV.  This should be regarded
as only a rough estimate, since
the physical value of $M_b$ is probably not large enough for
the constant $H_8'(\mu)$ to be negligible.

The subprocesses in the factorization formulas (\ref{facth}) and
(\ref{factchi})
are decays of the $b$ quark into $c {\bar c}$ plus other quarks and gluons.
The dominant contributions involve the coupling of the $b$ quark
to $c {\bar c} s$ via an effective 4-quark
weak interaction \cite{glam}, which can, by Fierz rearrangement,
be put into the form
$$
{\cal L}_{weak} \;=\; - {G_F \over \sqrt{2}} V_{cb} V_{cs}^*
\Bigg( {2 C_+ - C_- \over 3} \; {\bar c} \gamma_\mu (1 - \gamma_5) c
		\; {\bar s} \gamma^\mu (1 - \gamma_5) b
\quad \quad \quad \quad \quad \quad \quad
$$
\begin{equation} {
\quad \quad \quad \quad \quad \quad \quad
\;+\; (C_+ + C_-) \; {\bar c} \gamma_\mu (1 - \gamma_5) T^a c
		\; {\bar s} \gamma^\mu (1 - \gamma_5) T^a b \Bigg) \;,
} \label{Lweak}
\end{equation}
where $G_F$ is the Fermi constant and the $V_{ij}$'s are elements of the
Kobayashi-Maskawa mixing matrix.  The weak interaction that gives
the Cabibbo-suppressed transition $b \rightarrow c {\bar c} d$
is obtained by replacing ${\bar s}$ by ${\bar d}$
and $V_{cs}^*$ by $V_{cd}^*$ in (\ref{Lweak}).
The coefficients $C_+$ and $C_-$ in (\ref{Lweak}) are Wilson
coefficients that arise from evolving the effective 4-quark interaction
mediated by the $W$ boson from the scale $M_W$ down to the scale $M_b$.
To leading order in $\alpha_s(M_b)$
and to all orders in $\alpha_s(M_b) \ln(M_W/M_b)$, they are
\begin{equation} {
C_+(M_b) \;\approx\;
\left( {\alpha_s(M_b) \over \alpha_s(M_W)} \right)^{-6/23} \;,
} \label{cplus}
\end{equation}
\begin{equation}{
C_-(M_b) \;\approx\;
\left( {\alpha_s(M_b) \over \alpha_s(M_W)} \right)^{12/23} \;.
} \label{cminus}
\end{equation}
Taking $\alpha_s(M_W) = 0.116$ and $\alpha_s(M_b) = 0.20$,
we find that $C_+(M_b) \approx 0.87$
and  $C_-(M_b) \approx 1.34$.  When a $b$ quark decays through
the interaction (\ref{Lweak}), the first term
produces a $c {\bar c}$ pair in a color-singlet state,
while the second produces a $c {\bar c}$ pair in a color-octet state.
The color-singlet coefficient $2 C_+ - C_-$ is decreased dramatically
by renormalization-group evolution, from 1 at the scale $M_W$,
to $2 C_+ - C_- \approx 0.40$ at the scale $M_b$.
The color-octet coefficient $(C_+ + C_-)/2$
is increased slightly, from 1 at the scale $M_W$,
to $(C_+ + C_-)/2 \approx 1.10$ at the scale $M_b$.
Since the dramatic suppression of $2C_+ - C_-$ is due to a cancellation
between $2C_+$ and $C_-$, it is sensitive to both the choice $M_b$ for the
scale
and to higher-order perturbative corrections to the Wilson coefficients.
This sensitivity can be removed only by calculations beyond leading order.
The Wilson coefficients $C_+$ and $C_-$ have been calculated at
next-to-leading order \cite{acmp}, but the calculations are meaningful
only when combined with decay rates that are also calculated beyond
leading order.

It is convenient to express all the subprocess rates appearing in the
factorization formulas (\ref{factpsi}),
(\ref{facth}), and (\ref{factchi}) in terms of
\begin{equation} {
{\widehat \Gamma}_0 \;=\; |V_{cb}|^2 {G_F^2 \over 144 \pi} M_b^3 M_c
\left(1 - 4 {M_c^2 \over M_b^2} \right)^2 \; .
} \label{gamma}
\end{equation}
To leading order in $\alpha_s(M_b)$,
the color-singlet subprocess rates ${\widehat \Gamma}_1$
can be extracted from previous calculations \cite{mbw,dt,knr}.
The sum of the two subprocess rates that contribute to the
decay into the $\psi$ at leading order is
\begin{equation} {
{\widehat \Gamma}_1 \left( b \rightarrow c {\bar c} (^3S_1) + s, d \right)
\;=\; (2C_+ - C_-)^2
\left(1 + 8 {M_c^2 \over M_b^2} \right) \; {\widehat \Gamma}_0 \;,
} \label{gampsi}
\end{equation}
where we have used $|V_{cs}|^2 + |V_{cd}|^2 \approx 1$.
The subprocess rate that contributes to the decay into
$\chi_{c1}$ is \cite{knr}
\begin{equation} {
{\widehat \Gamma}_1 \left( b \rightarrow c {\bar c} (^3P_1) + s, d \right)
\;=\; 2 \; (2C_+ - C_-)^2
\left(1 + 8 {M_c^2 \over M_b^2} \right) \; {\widehat \Gamma}_0 \;.
} \label{gamchi}
\end{equation}
The color-singlet subprocess rates that contribute
to the production of $h_c$, $\chi_{c0}$, and $\chi_{c2}$ (the $^1P_1$,
$^3P_0$, and $^3P_2$ states, respectively)
vanish to leading order in $\alpha_s$ because of the
$J^{PC}$ quantum numbers of these charmonium states.

The  rate for direct production of $\psi$ in $B$ meson decay
has been calculated to next-to-leading order in $\alpha_s$ (Ref. \cite{chjc}).
Unfortunately, renormalization group effects were treated incorrectly
in that calculation.  The treatment of Ref. \cite{chjc} was equivalent to using
$2 C_+ - C_- = ( \alpha_s(M_b) / \alpha_s(M_W) )^{-24/23}$ and
$(C_+ + C_-)/2 = ( \alpha_s(M_b) / \alpha_s(M_W) )^{3/23}$
for the color-singlet and color-octet coefficients at the scale $M_b$.
This treatment correctly reproduces the term proportional to
$\alpha_s \ln(M_W/M_b)$ in the order $\alpha_s$ correction,
but it fails to reproduce the leading logarithms at order $\alpha_s^2$
and higher.  The results of Ref. \cite{chjc} were presented only
in graphical form, which prevents us from extracting the correct
order-$\alpha_s$ contribution to the subprocess rate for $\psi$ production.
In using the leading order result (\ref{gampsi}),
one should keep in mind that the next-to-leading correction
proportional to $\alpha_s (C_+ + C_-)^2$ may be as important numerically
as the leading term, which is proportional to $(2 C_+ - C_-)^2$.

The color-octet subprocess rates ${\widehat \Gamma}_8$
appearing in (\ref{facth}) and (\ref{factchi}) require a new calculation.
The most straightforward method (although not the simplest) is
to calculate the infrared divergent part of the rate for the decay
$b \rightarrow c {\bar c} s g$, with the $c {\bar c}$ in a color-singlet
P-wave state after having emitted the soft gluon.
This rate can be calculated in terms of
$R_P'(0)$ by making use of a covariant formalism \cite{kks}.
The infrared divergences come from the region of phase space
in which the momentum of the radiated gluon is soft.
In this region, the matrix element can be factored into
the amplitude for $b \rightarrow c {\bar c} s$, with the $c {\bar c}$ projected
onto an S state, and a term that depends on the gluon momentum.
The divergence comes from integrating over the phase space of the gluon.
Imposing an infrared cutoff $\mu$ on the energy of the soft gluon,
one finds that the divergence is proportional to $\ln(M_b/\mu) H_1$.
The final results for the infrared divergent terms in the decay rates are
\begin{equation} {
\Gamma(b \rightarrow h_c + s g)
\;\sim\; {12 \over \pi^2} |V_{cs}|^2 (C_+ + C_-)^2 \; \alpha_s
\ln{M_b \over \mu} \; {|R_P'(0)|^2 \over M_c^4} \; {\widehat \Gamma}_0 \;,
} \label{divh}
\end{equation}
\begin{equation} {
\Gamma(b \rightarrow \chi_{cJ} + s g)
\;\sim\; (2 J + 1) {4 \over 3 \pi^2} |V_{cs}|^2 (C_+ + C_-)^2 \; \alpha_s
\ln{M_b \over \mu} \; {|R_P'(0)|^2 \over M_c^4} \;
\left( 1 + 8 {M_c^2 \over M_b^2} \right) \; {\widehat \Gamma}_0 \;.
} \label{divchij}
\end{equation}
One can identify the corresponding infrared divergence in the
perturbative expression for $H_8'$
by neglecting the running of the coupling constant in
(\ref{rgeq}).  The resulting expression for $H_8'$ is
\begin{equation} {
H_8'(M_b) \; \sim \; {16 \over 27 \pi} \alpha_s
\ln \left( M_b \over \mu \right) \, H_1 \; .
} \label{hoctet}
\end{equation}
Using this identification together with the expression (\ref{hsinglet}) for
$H_1$, we find that the color-octet subprocess rates
defined in (\ref{facth}) and (\ref{factchi}) are
\begin{equation} {
{\widehat \Gamma}_8 \left( b \rightarrow c {\bar c}(^1S_0) + s, d \right)
\;=\; {3 \over 2} \;  (C_+ + C_-)^2 \; {\widehat \Gamma}_0 \;,
} \label{gamh}
\end{equation}
\begin{equation} {
{\widehat \Gamma}_8 \left( b \rightarrow c {\bar c}(^3S_1) + s, d \right)
\;=\; {1 \over 2} \; (C_+ + C_-)^2
\left(1 + 8 {M_c^2 \over M_b^2} \right) \; {\widehat \Gamma}_0 \;.
} \label{gamchij}
\end{equation}

We now turn to the phenomenological applications of our results.
Among the corrections to the factorization formulas for $B$ hadron decays
are the effects of spectator quarks and antiquarks,
which are suppressed by powers of $\Lambda/M_b$ (Ref. \cite{bigi}).
It is clear from decays of $D$ mesons that spectator effects
are much more important for total decay rates than for semileptonic decays.
Our predictions for the inclusive decay rates into charmonium states
should therefore be more reliable if they are normalized to
the semileptonic decay rate, instead of the total decay rate.
These branching ratios should be
identical for $B^-$, ${\bar B}^0$, ${\bar B}_s$, and $\Lambda_b$,
even if their lifetimes differ substantially.
To leading order in $\alpha_s$, the semileptonic decay rate is \cite{cm}
\begin{equation} {
\Gamma(b \rightarrow e^- {\bar \nu}_e + X)
\;=\; |V_{cb}|^2 {G_F^2 \over 192 \pi^3} M_b^5  \; F(M_c/M_b) \;,
} \label{Gamsl}
\end{equation}
where $F(x) = 1 - 8 x^2 - 24 x^4 \ln x + 8 x^6 - x^8$.
Approximating the ratio $M_c/M_b$ of the heavy quark masses by the ratio
$M_D/M_B \approx 0.35$ of the corresponding meson masses,
we find that the phase space factor is $F(M_c/M_b) \approx 0.41$
and that the semileptonic decay rate (\ref{Gamsl}) is
$|V_{cb}|^2 (M_b/5.3 {\rm GeV})^5 (3.9 \cdot 10^{-11} {\rm GeV})$.
The Kobayashi-Maskawa factor $|V_{cb}|^2$, as well as the extreme
sensitivity to the value of the bottom quark mass $M_b$,
cancels in the ratio between
the charmonium and semileptonic decay rates.

The leading order QCD prediction for the inclusive decay rate into $\psi$ is
obtained by inserting (\ref{gampsi}) into (\ref{factpsi}).  To take
into account phase space restrictions as accurately as possible,
we approximate the ratio $M_c/M_b$ of quark masses
by $M_{\psi}/(2 M_B) \approx 0.29$.  Normalizing to the semileptonic
decay rate, we obtain the ratio
\begin{equation} {
R(\psi) \;\equiv\;
{\Gamma(b \rightarrow \psi + X) \over
	\Gamma(b \rightarrow e^- {\bar \nu}_e + X)}
\;\approx\; 6.8 \; (2 C_+ - C_-)^2 {G_1 \over M_b} \;.
} \label{rpsi}
\end{equation}
Multiplying by the observed semileptonic branching fraction for $B$ mesons
of $10.7 \%$, we find that the predicted inclusive branching fraction for
$\psi$ is $0.23 \%$.  There are large theoretical uncertainties in this
result because order-$\alpha_s$ corrections to
the color-singlet Wilson coefficient $2C_+ - C_-$ may be significant
and because the perturbative corrections to the subprocess rate
(\ref{gampsi}) proportional to $\alpha_s (C_+ + C_-)^2$
need not be small compared to the leading-order
term, which is proportional to $(2C_+ - C_-)^2$.
These uncertainties can be removed only by a complete calculation of the
subprocess rate ${\widehat \Gamma}_1(b \rightarrow  \psi X)$
to order $\alpha_s$.  There are additional theoretical
errors due to uncertainties in the quark masses $M_c$ and $M_b$,
relativistic corrections of order $v^2$,
and corrections of order $\Lambda^2/M_b^2$.  However these
uncertainties are probably small compared to the perturbative errors.

For the P states, we define ratios
of the inclusive charmonium and semileptonic decay rates,
analogous to the ratio $R(\psi)$ defined in (\ref{rpsi}).
We take into account the phase space restrictions in the decay rate
into charmonium as accurately as possible
by using half the bound state mass for $M_c$
and the $B$ meson mass for $M_b$, so that $M_c/M_b$
varies from 0.32 to 0.34 for the various P states.
For the mass of the $h_c$, we have taken the
center of gravity of the $\chi_{cJ}$ states.
The resulting leading order QCD predictions for the ratios are then
\begin{equation} {
R(h_c) \;\approx\; 14.7 \; (C_+ + C_-)^2 {H_8'(M_b) \over M_b} \;,
} \label{rh}
\end{equation}
\begin{equation} {
R(\chi_{c0}) \;\approx\; 3.2 \; (C_+ + C_-)^2 {H_8'(M_b) \over M_b} \;,
} \label{rzero}
\end{equation}
\begin{equation} {
R(\chi_{c1}) \;\approx\; 12.4 \; (2 C_+ - C_-)^2 {H_1 \over M_b}
\;+\; 9.3 \; (C_+ + C_-)^2 {H_8'(M_b) \over M_b} \;,
} \label{rone}
\end{equation}
\begin{equation} {
R(\chi_{c2}) \;\approx\; 15.3 \; (C_+ + C_-)^2 {H_8'(M_b) \over M_b} \;.
} \label{rtwo}
\end{equation}
These predictions are subject to the same uncertainties as the prediction
for $\psi$ given in (\ref{rpsi}).

The ratios predicted above are for the direct production of
charmonium states in the decay of the $B$ hadron.
The branching fractions that are measured directly by experiment
include indirect production from cascade decays of higher charmonium states.
The branching fractions for direct production can only be obtained
by making assumptions about the cascade decays.
The inclusive branching fractions that have been measured are
$(1.12 \pm 0.16)\%$ for $\psi$ \cite{pdg},
$(0.46 \pm 0.20)\%$ for $\psi'$ \cite{pdg},
$(0.54 \pm 0.21)\%$ for $\chi_{c1}$ \cite{rap},
and an upper bound of $0.4 \%$ for $\chi_{c2}$ \cite{rap}.
The branching fractions for the cascade processes
$\psi' \rightarrow \psi + X$, $\psi' \rightarrow \chi_{c1} + \gamma$,
and $\chi_{c1} \rightarrow \psi + \gamma$ are
approximately $57 \%$, $9 \%$, and $27 \%$, respectively \cite{pdg}.
If one assumes that there are no other cascade processes that give
significant contributions, then the branching fractions for
direct production by $B$ meson decay
are $(0.71 \pm 0.20) \%$ for $\psi$,
$(0.46 \pm 0.20)\%$ for $\psi'$, and $(0.50 \pm 0.21)\%$ for $\chi_{c1}$.
The ratio of the direct production rates of
$\psi'$ and $\psi$ is consistent, within the large error bars,
with the ratio of their electronic widths, which is 0.40 \cite{pdg}.
However the result for $\psi$ is several standard deviations larger than the
prediction of (\ref{rpsi}). The discrepancy could be due to
order-$\alpha_s$ corrections to the Wilson coefficient $2C_+ - C_-$
or to corrections to the subprocess rate (\ref{gampsi})
proportional to $\alpha_s (C_+ + C_-)^2$.  Alternatively, it could be due to
contributions from cascade decays into $\psi$ and $\psi'$ from
higher charmonium states.  Studies of the momentum distribution
\cite{bklp} and the polarization \cite{mbw,knr} of the $\psi$'s could
help determine whether cascades from states other than $\psi'$
and $\chi_{c1}$ are important.

The measurement of the rate for $B$ meson decay
into $\chi_{c1}$ (Ref. \cite{rap}) makes it possible to
extract a phenomenological value for the unknown matrix element $H_8'$.
Dividing the branching fraction for direct decay of the $B$ meson
into $\chi_{c1}$ by
the semileptonic branching fraction $(10.7 \pm 0.5) \%$ \cite{pdg},
we find the branching ratio (\ref{rone}) to be $0.047 \pm 0.020$.
With the choice $M_b = 5.3$ GeV,
the color-singlet term on the right side of (\ref{rone}) is $0.006$.
Attributing the remainder to the color-octet term, we find that
$H_8'(M_b) \approx (4.8 \pm 2.3)$ MeV.  The quoted error is due to the
experimental error in the branching ratio only, and does not include the
theoretical error due to the potentially large perturbative corrections
to the Wilson coefficient $2C_+ - C_-$ and to the color-singlet
subprocess rate ${\widehat \Gamma}_1$.
One can also use (\ref{rtwo}) together with the upper
bound on the branching fraction into $\chi_{c2}$ to obtain the upper bound
$H_8'(M_b) < 2.7$ MeV.  This bound is consistent with the determination
from $\chi_{c1}$ production, given the large error bar.
The large theoretical uncertainty in our determination of $H_8'$
could be reduced dramatically by calculating the order-$\alpha_s$
corrections to the P-wave subprocess rates.  In the absence of
these calculations, one should regard our determination of $H_8'$
as only a rough estimate.

The color-octet mechanism has a dramatic effect on the pattern of
the production rates for P-wave charmonium states.
Taking the value $H_8'(M_b) \approx 2.5$ MeV, which is consistent with
both the determination from decays into $\chi_{c1}$ and
with the upper bound from decays into $\chi_{c2}$,
we find that the relative production rates
predicted by (\ref{rh})-(\ref{rtwo}) are
$h_c : \chi_{c0} : \chi_{c1} : \chi_{c2} \;=\; 1.3 : 0.3 : 1 : 1.3$.
A previous calculation of the
decay rates of $B$ mesons into charmonium states \cite{knr},
which did not take into account the color-octet mechanism,
gave the relative rates
$h_c : \chi_{c0} : \chi_{c1} : \chi_{c2} \;=\; 0 : 0 : 1 : 0$.
The observation of $\chi_{c2}$ production at
a rate comparable to that for $\chi_{c1}$ production would be
a dramatic confirmation of the color-octet production mechanism and
would provide a direct measurement of the matrix element $H_8'(M_b)$.

P-wave production provides unique information on the production of
heavy-quark pairs and on their binding into quarkonium.
In this paper we have outlined a consistent factorization formalism
for describing these processes in QCD.
In addition to the familiar color-singlet production mechanism,
the factorization formulas take into account the  color-octet mechanism,
in which a heavy quark-antiquark pair is created at short distances in a
color-octet S-wave state and subsequently fragments by a nonperturbative
process into a color-singlet P-wave bound state.  The
inclusive production rates for all four P states
can be calculated in terms of two nonperturbative inputs:
a parameter $H_1$ that is related to the derivative of the wavefunction
at the origin and a second parameter $H_8'$ that gives the
probability for the fragmentation from the color-octet S-wave state.
We have applied this formalism to the production
of P-wave charmonium in decays of $B$ hadrons
and have used experimental data to obtain a crude
estimate of the parameter $H_8'$.  An accurate determination of $H_8'$
requires perturbative calculations beyond leading order as well as
more accurate experimental data on $\chi_c$ production in $B$ meson decays.
The accurate determination of this parameter would be very useful because
the production rates for P-wave charmonium states in
photoproduction, leptoproduction,
hadron collisions, and other high energy physics processes
satisfy factorization formulas involving the same two nonperturbative
parameters $H_1$ and $H_8'$ that appear in $B$ meson decay.
Knowledge of these two parameters would, in principle, allow one to
compute all of the inclusive P-wave charmonium production rates.

This work was supported in part by the U.S. Department of Energy,
Division of High Energy Physics,
under Contract W-31-109-ENG-38 and under Grant DE-FG02-91-ER40684,
and by the National Science Foundation.
\vfill\eject

\vfill\eject

\end{document}